\documentclass[aps,twocolumn,showpacs,preprintnumbers,a4paper]{revtex4}
\usepackage{graphicx}
\newcommand{\nc}{\newcommand}
\nc{\rnc}{\renewcommand}
\nc{\beq}{\begin{equation}}
\nc{\eeq}{\end{equation}}
\nc{\bea}{\begin{eqnarray}}
\nc{\eea}{\end{eqnarray}}
\nc{\ba}{\begin{array}}
\nc{\ea}{\end{array}}
\nc{\nn}{\nonumber}
\nc{\bpi}{\begin{picture}}
\nc{\epi}{\end{picture}}
\nc{\scs}{\scriptstyle}
\nc{\unit}{{\mbox{\boldmath\large $1$}}}
\nc{\p}{\partial}
\nc{\ua}{\uparrow}
\nc{\da}{\downarrow}
\nc{\uada}{{\uparrow\downarrow}}

\nc{\al}{\alpha}
\nc{\be}{\beta}
\nc{\ga}{\gamma}
\nc{\de}{\delta}
\nc{\la}{\lambda}
\nc{\si}{\sigma}
\nc{\Ga}{{\sf\Gamma}}
\nc{\La}{\Lambda}

\nc{\abar}{\bar{a}}
\nc{\J}{{\sf J}}
\nc{\T}{{\sf T}}
\rnc{\P}{{\sf P}}
\nc{\R}{{\sf R}}
\rnc{\S}{{\sf S}}
\nc{\U}{{\sf U}}
\nc{\V}{{\sf V}}
\nc{\Va}{{\sf V}_a}
\nc{\Vap}{{\sf V}_a'}
\nc{\Vb}{{\sf V}_b}
\nc{\Vc}{{\sf V}_c}
\nc{\Vd}{{\sf V}_d}
\nc{\X}{{\sf X}}
\nc{\Y}{{\sf Y}}
\nc{\Z}{{\sf Z}}

\nc{\cd}[2]{\scs\{\raisebox{-.46ex}{\rlap{\tiny{#2}}}
  \raisebox{.45ex}{\tiny{#1}}\scs\}}
\nc{\zm}{\ba{cc}0&0\\0&0\ea}
\nc{\zr}{\ba{cc}0&0\ea}
\nc{\zc}{\ba{c}0\\0\ea}

\begin{document}

\title{Pedestrian Solution of the Two-Dimensional Ising Model}
%=============================================================
\author{Boris Kastening}
\email[Email address: ]{ka@physik.fu-berlin.de}
\affiliation{Institut f\"ur Theoretische Physik\\
Freie Universit\"at Berlin\\
Arnimallee 14\\
D-14195 Berlin\\
Germany}

\date{20 April 2001}

\begin{abstract}
The partition function of the two-dimensional Ising model with zero
magnetic field on a square lattice with $m\times n$ sites wrapped on a
torus is computed within the transfer matrix formalism in an explicit
step-by-step approach inspired by Kaufman's work.
However, working with two commuting representations of the complex
rotation group SO$(2n,C)$ helps us avoid a number of unnecessary
complications.
We find all eigenvalues of the transfer matrix and therefore the
partition function in a straightforward way.
\end{abstract}


\maketitle

\section{Introduction}
%---------------------
Since Onsager's solution \cite{onsager} in the transfer matrix approach
\cite{kw} of the two-dimensional Ising model \cite{li} with vanishing
magnetic field and its subsequent simplification by Kaufman \cite{kaufman},
there have been a number of related as well as alternative solutions, see
e.g.\ Baxter's book on exactly solved models in statistical mechanics
\cite{baxterbook} and references therein.
Among the transfer matrix solutions are the ones by Schultz, Mattis
and Lieb, by Thompson, by Baxter and by Stephen and Mittag \cite{tmf}.
Nevertheless, the author of this work feels that there is still room for
a nice and straightforward solution.
A much abbreviated version of this work, sketching only the new aspects
may be found in \cite{bo}, while here we give a self-contained account
of our approach with all relevant details.

We study the two-dimensional Ising model with zero magnetic field on a
square lattice with $m$ rows and $n$ columns subject to toroidal
boundary conditions.
The transfer matrix is expressed in terms of the generators of two
commuting representations of the complex rotation group SO$(2n,C)$.
These representations naturally arise from projected bilinears of
$2^n\times2^n$ spin matrices.
Conservatively speaking, we reduce Kaufman's approach to its essential
steps, avoiding in particular the doubling of the number of eigenvalues
of the transfer matrix and subsequent rather involved arguments for the
choice of the correct ones.
We also need not investigate the transformation properties of the spin
matrices.
In our opinion, the value of this work is not that it gives the most
concise solution, but that it provides a clear line of arguments and
requires only rather basic tools from mathematics and mathematical physics.

We proceed in a step-by-step fashion, exhibiting also those steps
that closely parallel the corresponding steps in Kaufman's treatment
\cite{kaufman} or Huang's corresponding textbook write up \cite{huang}.
On the one hand this is necessary because of our slightly different (but,
in our opinion, more natural) conventions, in particular for $\ga_0$.
On the other hand, we would like to keep this work as self-contained as
possible.

Finally, this work may help sharpen the view on the difficulties
encountered when attempting to solve the two-dimensional model with
magnetic field or the three-dimensional model and might be useful for
simplifying other problems in statistical mechanics for which an exact
solution is attempted.

Our notation is in the spirit of \cite{huang}, with sans serif capitals
reserved for $2^n\times2^n$ matrices.
The structure of this work is as follows:
In section \ref{def} the model and its transfer matrix $\T$ are defined.
In section \ref{xyz} we express the transfer matrix in
terms of $2^n\times2^n$ spin matrices $\X_\nu$, $\Y_\nu$, $\Z_\nu$.
A rescaled transfer matrix $\V$ is defined whose eigenvalues are, up to a
trivial factor, the eigenvalues of $\T$.
In section \ref{gammaandj}, we define further spin matrices $\Ga_\nu$
and two commuting projection classes $\J_{\al\be}^+$ and $\J_{\al\be}^-$
of their bilinears.
After investigating the relevant properties of the $\J_{\al\be}^\pm$,
we express $\V$ in terms of them.
In section \ref{vpmdiag}, we introduce $2n\times2n$ matrices $J_{\al\be}$
whose algebra is identical to that of $\J_{\al\be}^\pm$ and define matrices
$V^\pm$ in terms of the $J_{\al\be}$ such that the relation between $V^\pm$
and $J_{\al\be}$ is closely related to that between $\V$ and
$\J_{\al\be}^\pm$.
The $V^\pm$ are subsequently diagonalized.
In section \ref{vdiag}, the analogy between $V^\pm$ and $\V$ is exploited
for the diagonalization of $\V$.
The eigenvalues of $\V$ and the partition function are found explicitly.
In section \ref{extensions} we comment on the technical difficulties
encountered when attempting to solve the two-dimensional model with
magnetic field or the three-dimensional model.

\section{The Model and its Transfer Matrix \boldmath$\T$}
%--------------------------------------------------------
\label{def}
We work with a square lattice with $m$ rows and $n$ columns and
consequently $m\times n$ sites.
\begin{figure}[t]
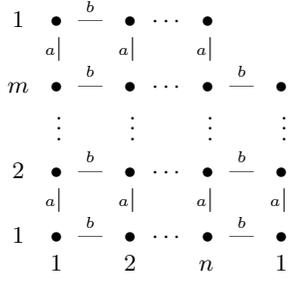

\begin{center}
$\ba{crcrcrcr}
1 & \bullet & \stackrel{b}{\mbox{---}} & \bullet & \cdots & \bullet \\
& {\scs a}| && {\scs a}| && {\scs a}| \\
m & \bullet & \stackrel{b}{\mbox{---}} & \bullet & \cdots & \bullet &
\stackrel{b}{\mbox{---}} & \bullet \\
& \vdots && \vdots && \vdots && \vdots \\
2 & \bullet & \stackrel{b}{\mbox{---}} & \bullet & \cdots & \bullet &
\stackrel{b}{\mbox{---}} & \bullet \\
& {\scs a}| && {\scs a}| && {\scs a}| && {\scs a}| \\
1 & \bullet & \stackrel{b}{\mbox{---}} & \bullet & \cdots & \bullet &
\stackrel{b}{\mbox{---}} & \bullet \\
& 1 && 2 && n && 1\\
\ea
$
\end{center}
\caption{\label{twodimlattice}
Lattice of the two-dimensional Ising model.
The parameters $a$ and $b$ correspond to the interactions in the
respective direction indicated.
Rows $1$ and $m{+}1$ and columns $1$ and $n{+}1$ are identified,
respectively, i.e.\ the model is wrapped on a torus.}
\end{figure}
The energy is given by
\beq
E=E_a+E_b,
\eeq
\bea
-\be E_a
&=&
a\sum_{\langle ij\rangle_a}s_i s_j
=a\sum_{\mu=1}^m\sum_{\nu=1}^ns_{\mu\nu}s_{\mu+1,\nu},
\eea
\bea
-\be E_b
&=&
b\sum_{\langle ij\rangle_b}s_i s_j
=b\sum_{\mu=1}^m\sum_{\nu=1}^ns_{\mu\nu}s_{\mu,\nu+1},
\eea
with $a=-\be J_a$, $b=-\be J_b$ and $\be^{-1}=kT$, where
$J_a$ and $J_b$ are temperature-independent interaction energy
parameters which by convention are negative (other sign combinations
do not change the partition function, as can be seen by appropriate
redefinitions of half of the signs of the spin variables).
Consequently, $a$ and $b$ are real and positive.
$\langle ij\rangle_a$ and $\langle ij\rangle_b$ indicate summation
over neighboring spins in the respective directions.
We identify rows $1$ and $m{+}1$ and columns $1$ and $n{+}1$, i.e.\
the lattice is wrapped on a torus, see also Fig.~\ref{twodimlattice}.
The $s_{\mu\nu}$ can take the values $\pm1$.
The partition function is then given by
\bea
\label{psum}
\lefteqn{Z(a,b)=\sum_{s_{11}}\cdots\sum_{s_{mn}}\exp(-\be E)}
\nn\\
&=&
\sum_{s_{11}}\cdots\sum_{s_{mn}}
\exp\left[\sum_{\mu=1}^m\sum_{\nu=1}^n\left(as_{\mu\nu}s_{\mu+1,\nu}
+bs_{\mu\nu}s_{\mu,\nu+1}\right)\right]
\nn\\
&=&
\sum_{s_{11}}\cdots\sum_{s_{mn}}\prod_{\mu=1}^m\prod_{\nu=1}^n
\exp(as_{\mu\nu}s_{\mu+1,\nu}+bs_{\mu\nu}s_{\mu,\nu+1}).
\eea
It can be expressed with the help of a $2^n\times2^n$ transfer matrix $\T$,
\bea
\lefteqn{Z(a,b)}
\nn\\
&=&
\sum_{\{\pi_i\}}
\langle\pi_1|\T|\pi_m\rangle
\langle\pi_m|\T|\pi_{m-1}\rangle\cdots
\langle\pi_3|\T|\pi_2\rangle
\langle\pi_2|\T|\pi_1\rangle
\nn\\
&=&
\mbox{Tr}\,\T^m,
\eea
where $\pi_\mu=\{s_{\mu1},\ldots,s_{\mu n}\}$ for $\mu=1,\ldots,m$
and $\T$ is defined by its elements ($s_{n+1}\equiv s_1$),
\beq
\langle\pi|\T|\pi'\rangle
=\prod_{\nu=1}^n\exp(as_\nu s_\nu'+bs_\nu s_{\nu+1}).
\eeq
We can split $\T$ into a product of two matrices
\beq
\label{t}
\T=\Vb\Vap,
\eeq
defining $\Va'$ and $\Vb$ by their elements 
\beq
\langle\pi|\Vap|\pi'\rangle
=\prod_{\nu=1}^n\exp(as_\nu s_\nu')
\eeq
and
\beq
\langle\pi|\Vb|\pi'\rangle
=\prod_{\nu=1}^n
\de_{s_\nu s_\nu'}\exp(bs_\nu s_{\nu+1}).
\eeq

\section{\boldmath$\T$ and Spin Matrices $\X_\nu$, $\Y_\nu$, $\Z_\nu$}
%---------------------------------------------------------------------
\label{xyz}
With the help of the Pauli matrices
\beq
\si_x=\left(\ba{rr}&1\\1\ea\right),\;\;\;\;\;\;
\si_y=\left(\ba{rr}&-i\\i\ea\right),\;\;\;\;\;\;
\si_z=\left(\ba{rr}1\\&-1\ea\right)
\eeq
and the $2\times2$ unit matrix $\unit$, define Hermitian $2^n\times2^n$
spin matrices by the direct products
\beq
\X_\nu=
\underbrace{\unit\otimes\cdots\otimes\unit}_{\nu-1}\otimes\;\si_x\otimes
\underbrace{\unit\otimes\cdots\otimes\unit}_{n-\nu},
\eeq
and analogously for $\Y_\nu$ and $\Z_\nu$.
For any $\nu$ and $\nu'$, they obey the commutation relations
\beq
[\X_\nu,\X_{\nu'}]=[\Y_\nu,\Y_{\nu'}]
=[\Z_\nu,\Z_{\nu'}]=0,
\eeq
while only for $\nu\neq\nu'$ holds
\beq
[\X_\nu,\Y_{\nu'}]=[\Y_\nu,\Z_{\nu'}]
=[\Z_\nu,\X_{\nu'}]=0.
\eeq
For any $\nu$ we have
\beq
\X_\nu^2=\Y_\nu^2=\Z_\nu^2=\unit,
\eeq
\beq
\{\X_\nu,\Y_\nu\}=\{\Y_\nu,\Z_\nu\}=\{\Z_\nu,\X_\nu\}=0,
\eeq
\beq
\X_\nu\Y_\nu=i\Z_\nu,\;\;\;\;\;\;
\Y_\nu\Z_\nu=i\X_\nu,\;\;\;\;\;\;
\Z_\nu\X_\nu=i\Y_\nu,
\eeq
where $\unit$ is now the $2^n\times2^n$ unit matrix.

Define $\abar$ by
\beq
\sinh(2\abar)\sinh(2a)=1,
\eeq
so that $\abar>0$,
\beq
\tanh\abar=\exp(-2a),\;\;\;\;\;\;\tanh a=\exp(-2\abar)
\eeq
and
\beq
\left(\ba{ll}e^{+a}&e^{-a}\\e^{-a}&e^{+a}\ea\right)
=[2\sinh(2a)]^{1/2}\exp(\abar\si_x).
\eeq
Then we can write
\beq
\Vap=[2\sinh(2a)]^{n/2}\Va
\eeq
with
\beq
\label{va}
\Va=\prod_{\nu=1}^n\exp\left(\abar\X_\nu\right),
\eeq
and
\beq
\label{vb}
\Vb=\prod_{\nu=1}^n\exp\left(b\Z_\nu\Z_{\nu+1}\right),
\eeq
where we have identified $\Z_{n+1}=\Z_1$.
The transfer matrix (\ref{t}) may then be expressed as
\beq
\T=[2\sinh(2a)]^{n/2}\Vb\Va.
\eeq
Due to the cyclic property of the trace, we may rewrite the partition
function (\ref{psum}) as
\beq
Z(a,b)=[2\sinh(2a)]^{mn/2}\mbox{Tr}\V^m,
\eeq
where $\V$ is defined by the Hermitian matrix
\beq
\label{v}
\V=\V_{a/2}\Vb\V_{a/2}
\eeq
with
\beq
\label{vahalf}
\V_{a/2}=\prod_{\nu=1}^n\exp(\abar\X_\nu/2),
\eeq
so that $\V_{a/2}^2=\Va$.
If $\La_k$ are the $2^n$ eigenvalues of $\V$, we have
\beq
Z(a,b)=[2\sinh(2a)]^{mn/2}\sum_{k=1}^{2^n}\La_k^m.
\eeq
Our task is therefore to find the eigenvalues of $\V$.

\section{Spin Matrices \boldmath$\Ga_\nu$ and Algebras of
Their Projected Bilinears $\J_{\al\be}^\pm$}
%--------------------------------------------------------
\label{gammaandj}
Define the $2n$ matrices ($\nu=1,\ldots,n$)
\bea
\Ga_{2\nu-1}
&=&
\X_1\cdots \X_{\nu-1}\Z_\nu,\\
\Ga_{2\nu}
&=&
\X_1\cdots \X_{\nu-1}\Y_\nu,
\eea
which obey
\beq
\label{gaga}
\{\Ga_\mu,\Ga_\nu\}=2\de_{\mu\nu}.
\eeq
Define further the matrix
\beq
\label{ux}
\U_\X=\X_1\cdots \X_n=i^n\Ga_1\Ga_2\cdots\Ga_{2n},~~~~~~\U_\X^2=\unit,
\eeq
which anticommutes with every $\Ga_\mu$,
\beq
\{\Ga_\mu,\U_\X\}=0.
\eeq
Now we can write for the matrices appearing in the exponents of
(\ref{vb}) and (\ref{vahalf})
\bea
\label{x}
\X_\nu&=&-\frac{i}{2}[\Ga_{2\nu},\Ga_{2\nu-1}],~~~~~\nu=1,\ldots,n,\\
\label{zz1}
\Z_\nu\Z_{\nu+1}&=&-\frac{i}{2}[\Ga_{2\nu+1},\Ga_{2\nu}],
~~~~~\nu=1,\ldots,n-1,\\
\label{zz2}
\Z_n\Z_1&=&\frac{i}{2}\U_\X[\Ga_1,\Ga_{2n}].
\eea

So far our treatment has been rather similar to Huang's write up
\cite{huang} of Kaufman's approach \cite{kaufman}.
Our subsequent treatment rests on the observation that the formulation
(\ref{v}) of $\V$ with (\ref{vb}) and (\ref{vahalf}) involves only
the product $\U_\X$ of all $\Ga_\nu$ and bilinears $\Ga_\al\Ga_\be$,
see (\ref{ux}), (\ref{x})-(\ref{zz2}).
This will allow us to find two commuting algebras of projected bilinears
of the $\Ga_\nu$.

Define the projectors
\beq
\P^\pm\equiv\frac{1}{2}(\unit\pm\U_\X),
\eeq
for which hold
$$
\P^++\P^-=\unit,~~~~
\P^+\P^+=\P^+,~~~~\P^-\P^-=\P^-,
\vspace{-20pt}
$$
\beq
\vspace{-20pt}
\eeq
$$
\P^+\P^-=\P^-\P^+=0,~~~~
\P^\pm\U_\X=\U_\X\P^\pm=\pm\P^\pm
$$
and
\beq
[\P^\pm,\Ga_\al\Ga_\be]=0.
\eeq
With their help, define the matrices
\beq
\label{jgaga}
\J_{\al\be}=-\frac{i}{4}[\Ga_\al,\Ga_\be],~~~~~~
\J_{\al\be}^\pm=\P^\pm\J_{\al\be},~~~~~~
\eeq
so that
\beq
\label{jprops}
\J_{\al\be}=\J_{\al\be}^++\J_{\al\be}^-,~~~~~~
\U_\X\J_{\al\be}^\pm=\pm\J_{\al\be}^\pm.
\eeq
Since
\beq
\J_{\al\be}^\pm=-\J_{\be\al}^\pm,
\eeq
there are $n(2n-1)$ such independent matrices of each 
kind $\J_{\al\be}^+$ and $\J_{\al\be}^-$.
It is straightforward to show that their algebra decomposes into
two commuting parts,
\beq
[\J_{\al\be}^+,\J_{\ga\de}^-]=0,
\eeq
which obey identical algebras
\beq
\label{jalgebra}
[\J_{\al\be}^\pm,\J_{\ga\de}^\pm]
=i(\de_{\al\ga}\J_{\be\de}^\pm+\de_{\be\de}\J_{\al\ga}^\pm
-\de_{\al\de}\J_{\be\ga}^\pm-\de_{\be\ga}\J_{\al\de}^\pm).
\eeq

Next note that with (\ref{x})-(\ref{zz2}), (\ref{jgaga}) and (\ref{jprops})
we can write
\bea
\X_\nu&=&2(\J_{2\nu,2\nu-1}^+{+}\J_{2\nu,2\nu-1}^-),
~~~~\nu=1,\ldots,n,
\\
\Z_\nu\Z_{\nu+1}&=&2(\J_{2\nu+1,2\nu}^+{+}\J_{2\nu+1,2\nu}^-),
~~~~\nu=1,\ldots,n{-}1,\;\;\;\;\;\;
\\
\Z_n\Z_1&=&-2\U_\X(\J_{1,2n}^+{+}\J_{1,2n}^-)
=-2(\J_{1,2n}^+{-}\J_{1,2n}^-).
\eea
This allows us to express $\V_{a/2}$ from (\ref{vahalf}) and $\Vb$ from
(\ref{vb}) in terms of the $\J_{\al\be}^\pm$,
\beq
\V_{a/2}=\prod_{\nu=1}^n\exp[\abar(\J_{2\nu,2\nu-1}^++\J_{2\nu,2\nu-1}^-)]
=\V_{a/2}^+\V_{a/2}^-
\eeq
with
\beq
\label{vapm}
\V_{a/2}^\pm=\prod_{\nu=1}^n\exp(\abar\J_{2\nu,2\nu-1}^\pm),
\eeq
and
\bea
\label{vb3}
\Vb
&=&
\exp[-2b(\J_{1,2n}^+-\J_{1,2n}^-)]
\nn\\
&&\times
\prod_{\nu=1}^{n-1}\exp[2b(\J_{2\nu+1,2\nu}^++\J_{2\nu+1,2\nu}^-)]
\nn\\
&=&
\Vb^+\Vb^-
\eea
with
\beq
\label{vbpm}
\Vb^\pm=\exp(\mp2b\J_{1,2n}^\pm)
\prod_{\nu=1}^{n-1}\exp(2b\J_{2\nu+1,2\nu}^\pm).
\eeq
The rescaled transfer matrix $\V$ defined in (\ref{v}) reads then
\beq
\V=\V^+\V^-
\eeq
with
\beq
\label{tvv}
\V^\pm=\V_{a/2}^\pm\Vb^\pm\V_{a/2}^\pm,~~~~~~[\V^+,\V^-]=0.
\eeq

\section{The Matrices \boldmath$V^\pm$ and Their Diagonalization}
%----------------------------------------------------------------
\label{vpmdiag}
In this section we formulate matrices $V^{\pm}$ whose definitions are
similar to the expressions of $\V^\pm$ through (\ref{vapm}), (\ref{vbpm})
and (\ref{tvv}).
However, the $V^\pm$ have convenient periodicity properties which allow
for explicit diagonalization.
Our treatment closely parallels that of Kaufman \cite{kaufman} as
written up by Huang \cite{huang}.
However, we will be more explicit and also carefully analyze some
subtleties of the analysis concerning the representation of the
relevant similarity transformation matrices $S_\pm$ and the sign of
$\ga_0$, see below.

Define $N\times N$ matrices $J_{\al\be}$ by their elements
\beq
\left(J_{\al\be}\right)_{ij}
=-i(\de_{\al i}\de_{\be j}-\de_{\be i}\de_{\al j}),
\eeq
where Greek and Latin indices run from 1 to $N$.
Since $J_{\al\be}=-J_{\be\al}$, there are $N(N-1)/2$ such independent
matrices.
As can be easily checked, they obey the algebra
\beq
\label{sonalgebra}
[J_{\al\be},J_{\ga\de}]=i(\de_{\al\ga}J_{\be\de}+\de_{\be\de}J_{\al\ga}
-\de_{\al\de}J_{\be\ga}-\de_{\be\ga}J_{\al\de}),
\eeq
which is, if we set $N=2n$, identical to the algebras (\ref{jalgebra})
of $\J_{\al\be}^+$ and $\J_{\al\be}^-$.
Now consider the matrices
\beq
S=\exp(ic_{\al\be}J_{\al\be}),
\eeq
for which holds
\bea
S^T
&=&
\exp(ic_{\al\be}J_{\al\be})^T=\exp(ic_{\al\be}J_{\al\be}^T)
\nn\\
&=&
\exp(-ic_{\al\be}J_{\al\be})=S^{-1}.
\eea
Since
\beq
(\det S)^2=\det S\det S^T=\det S\det S^{-1}=1,
\eeq
and because $S$ is smoothly connected to the unit matrix, we have
$\det S=1$.
For real parameters $c_{\al\be}$, the $S$ are also real and form the
group SO$(N)$ of $N\times N$ orthogonal matrices with unit determinant.
The algebra (\ref{sonalgebra}) is therefore called the Lie algebra
of SO$(N)$.
Here, we let the $c_{\al\be}$ be arbitrary complex numbers, so the
matrices $S$ form the group SO$(N,C)$ of complex $N\times N$
matrices with
\beq
S^T=S^{-1},~~~~~~\det S=1.
\eeq

Define the SO$(2n,C)$ matrices
\beq
\label{vpm}
V^\pm=V_{a/2}V_b^\pm V_{a/2}
\eeq
with
\beq
V_{a/2}=\prod_{\nu=1}^n\exp(\abar J_{2\nu,2\nu-1})
\eeq
and
\beq
V_b^\pm=\exp(\mp2b J_{1,2n})
\prod_{\nu=1}^{n-1}\exp(2b J_{2\nu+1,2\nu}),
\eeq
in analogy with $\V^\pm$, $\V_{a/2}^\pm$ and $\Vb^\pm$ in (\ref{tvv}),
(\ref{vapm}) and (\ref{vbpm}).
Since $\abar$ and $b$ are real, the matrices $V_{a/2}$, $V_b^\pm$ and
$V^\pm$ are not only orthogonal, but also Hermitian, so the $V^\pm$ have only
real eigenvalues and in each case a complete set of orthonormal eigenvectors.
Since the algebras of the three sets of matrices $\J_{\al\be}^+$,
$\J_{\al\be}^-$ and $J_{\al\be}$ are identical and because $V^\pm$ will
turn out to be explicitly diagonalizable, we will be able to find all
eigenvalues of $\V$.

The structure of $V_{a/2}$ and $V_b^\pm$ is given by
\beq
V_{a/2}=\left(\ba{cccc}
\framebox{$R_{\abar}$}&\zm&\cdots&\zm\\
\zm&\framebox{$R_{\abar}$}&\cdots&\zm\\
\vdots&\vdots&\ddots&\vdots\\
\zm&\zm&\cdots&\framebox{$R_{\abar}$}
\ea\right)
\eeq
with
\beq
\framebox{$R_{\abar}$}=\left(\ba{rr}\cosh\abar&i\sinh\abar\\
-i\sinh\abar&\cosh\abar\ea\right)
\eeq
and
\beq
V_b^\pm=\left(\ba{ccccc}
\cosh(2b)&\zr&\cdots&\zr&\pm i\sinh2b\\
\zc&\framebox{$R_{2b}$}&\cdots&\zm&\zc\\
\vdots&\vdots&\ddots&\vdots&\vdots\\
\zc&\zm&\cdots&\framebox{$R_{2b}$}&\zc\\
\mp i\sinh2b&\zr&\cdots&\zr&\cosh(2b)
\ea\right)
\eeq
with
\beq
\framebox{$R_{2b}$}=\left(\ba{rr}\cosh2b&i\sinh2b\\
-i\sinh2b&\cosh2b\ea\right).
\eeq
Carrying out the matrix multiplication in (\ref{vpm}) is straightforward
and gives
\beq
V^\pm=\left(\ba{ccccccc}
A&B&0&\cdots&0&\mp B^\dag\\
B^\dag&A&B&\cdots&0&0\\
0&B^\dag&A&\cdots&0&0\\
\vdots&\vdots&\vdots&\ddots&\vdots&\vdots\\
0&0&0&\cdots&A&B\\
\mp B&0&0&\cdots&B^\dag&A
\ea\right)
\eeq
with
\beq
A=\cosh2b\left(\ba{rr}\cosh2\abar&i\sinh2\abar\\
-i\sinh2\abar&\cosh2\abar\ea\right)
\eeq
and
\beq
B=\sinh2b\left(\ba{rr}-\frac{1}{2}\sinh2\abar&-i\sinh^2\abar\\
i\cosh^2\abar&-\frac{1}{2}\sinh2\abar\ea\right).
\eeq

Let us make the following ansatz for the normalized eigenvectors of
$V^\pm$,
\beq
\label{psi}
\psi=\frac{1}{\sqrt{n}}\left(\ba{c}zu\\z^2u\\\vdots\\z^nu\ea\right),
\eeq
where $z$ is a complex number and $u$ is a two-component vector that we
assume to be normalized, $u^\dag u=1$.
The condition
\beq
V^\pm\psi=\la\psi
\eeq
leads to the $n$ equations
\bea
(zA+z^2B\mp z^nB^\dag)u&=&z\la u,\\
(z^2A+z^3B+zB^\dag)u&=&z^2\la u,\\
(z^3A+z^4B+z^2B^\dag)u&=&z^3\la u,\\
&\vdots&\nn\\
(z^{n-1}A+z^nB+z^{n-2}B^\dag)u&=&z^{n-1}\la u,\\
(z^nA\mp zB+z^{n-1}B^\dag)u&=&z^n\la u.
\eea
The equations with $z^2$ through $z^{n-1}$ on the right hand side
are identical, which leaves three independent equations
\bea
\label{abbeq1}
(A+zB\mp z^{n-1}B^\dag)u&=&\la u,\\
(A+zB+z^{-1}B^\dag)u&=&\la u,\\
\label{abbeq3}
(A\mp z^{1-n}B+z^{-1}B^\dag)u&=&\la u.
\eea
These equations can simultaneously be solved when
\beq
\label{z}
z^n=\mp1.
\eeq
since then they become identical,
\beq
\label{laeq}
(A+zB+z^{-1}B^\dag)u=\la u.
\eeq
The sign $\mp$ in (\ref{z}) is associated with $V^\pm$.
Altogether, there are $2n$ values of $z$ that solve (\ref{z}),
\beq
\label{zk}
z_k=e^{i\pi k/n},~~~~~~k=0,\ldots,2n{-}1.
\eeq
Even $k$ lead to $z^n=+1$ while odd $k$ lead to $z^n=-1$, i.e.\
\bea
k=1,3,5,\ldots,2n-1&&~~~\mbox{for $V^+$},\\
k=0,2,4,\ldots,2n-2&&~~~\mbox{for $V^-$}.
\eea
For each $k=0,\ldots,2n-1$, we still have to find the associated two
eigenvalues $\la_k^\ua$ and $\la_k^\da$ and their corresponding
eigenvectors $u_k^\ua$ and $u_k^\da$ in (\ref{laeq}),
i.e.\ in
\beq
\label{evecs}
M_ku_k^\uada=\la_k^\uada u_k^\uada
\eeq
with the Hermitian matrix
\beq
M_k=A+e^{i\pi k/n}B+e^{-i\pi k/n}B^\dag
=\left(\ba{cc}d_k&o_k\\o_k^*&d_k\ea\right)
\eeq
with
\beq
\label{dk1}
d_k=\cosh2\abar\cosh2b-\cos\frac{\pi k}{n}\sinh2\abar\sinh2b
\eeq
and
\bea
\label{ok1}
\lefteqn{o_k
=
-\sin\frac{\pi k}{n}\sinh2b}
\nn\\
&&{}
+i\left(\sinh2\abar\cosh2b
-\cos\frac{\pi k}{n}\cosh2\abar\sinh2b\right).\;\;\;\;
%\;\;\;\;
\eea
Explicit evaluation gives
\beq
\label{detm}
\det M_k=d_k^2-o_ko_k^*=1,
\eeq
implying [up to an overall sign, fixed by
${\rm tr}M_k=\la_k^\ua+\la_k^\da=2d_k>0$,
see (\ref{dk1}) above] eigenvalues of the form
\beq
\label{lak}
\la_k^\ua=e^{+\ga_k},~~~~~~
\la_k^\da=e^{-\ga_k}
\eeq
with real $\ga_k$.
The value of $\ga_k$ can be found from
\beq
{\rm tr}M_k=e^{+\ga_k}+e^{-\ga_k}
=2\cosh\ga_k.
\eeq
so that
\beq
\label{gak}
\cosh\ga_k=\cosh2\abar\cosh2b-\cos\frac{\pi k}{n}\sinh2\abar\sinh2b,
\eeq
and from (\ref{detm}) follows then that we may write
\beq
\label{ok2}
o_k=ie^{i\de_k}\sinh\ga_k
\eeq
with some phase $\de_k$ to be considered further below.

If $\ga_k$ is a solution, then also $-\ga_k$ is one, but this has already
been taken into account in (\ref{lak}).
Let us fix the sign of $\ga_k$ by defining $\ga_k=2\abar$ for $b=0$
and then analytically continuing to other values of $b$.
For $k=1,\ldots,2n-1$, this means $\ga_k>0$.
On the other hand, for $\ga_0$ this means
\beq
\label{ga0}
\ga_0=2(\abar-b).
\eeq
Our sign convention for the $\ga_k$ and in particular for $\ga_0$ will
allow us to treat all $\ga_k$ on an equal footing for both $\abar>b$ and
$\abar<b$, i.e.\ irrespective of the temperature.

\begin{figure}
\includegraphics[width=8cm,angle=0]{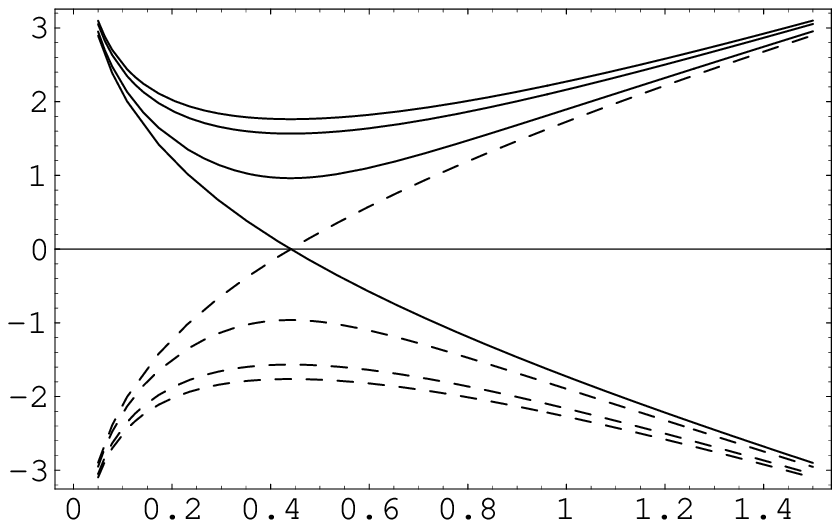}
\includegraphics[width=8cm,angle=0]{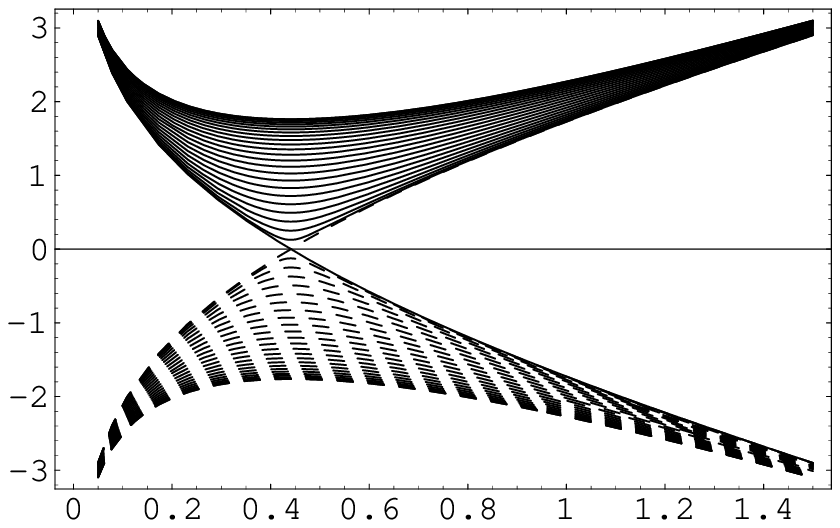}
\caption{\label{gakfig}
$\ga_k$ for $n=3$ and $n=25$ as a function of $a$
in the model with $b=a$.
The solid lines from the top are $\ga_n$, $\ga_{n-1}$, \ldots, $\ga_0$,
while the dashed lines from the bottom are $-\ga_n$, $-\ga_{n-1}$,
$\ldots$, $-\ga_0$.
Apart from $\pm\ga_0$ and $\pm\ga_n$, all $\pm\ga_k$ are two-fold degenerate.
For $n\rightarrow\infty$, the lines merge into two continua above and below
$0$.}
\end{figure}

It is obvious from (\ref{dk1}) that for $k=1,\ldots,n-1$
\beq
\label{gaksym}
\ga_k=\ga_{2n-k},
\eeq
i.e.\ the eigenvalues $\la_k^\uada$ with $k=1,\ldots,n-1$
occur in pairs.
Since for $0<k<n$ [extending for the moment (\ref{gak}) to non-integer
values of $k$],
\beq
\frac{\p\ga_k}{\p k}
=\frac{\pi\sinh2\abar\sinh2b}{n\sinh\ga_k}\sin\frac{\pi k}{n}>0,
\eeq
we have
\beq
0<|\ga_0|<\ga_1<\cdots<\ga_n.
\eeq
This means
\bea
0<+\ga_0<\ga_1<\cdots<\ga_n,~~~~~~\abar>b,
\\
0<-\ga_0<\ga_1<\cdots<\ga_n,~~~~~~\abar<b.
\eea
We have plotted two examples with $a=b$ in Fig.~\ref{gakfig}.

Comparison of (\ref{ok1}) with (\ref{ok2}) gives
\bea
\label{dek1}
\cos\de_k\sinh\ga_k
&\!=\!&
\sinh2\abar\cosh2b-\cos\frac{\pi k}{n}\cosh2\abar\sinh2b,
\nn\\\\
\label{dek2}
\sin\de_k\sinh\ga_k
&\!=\!&
\sin\frac{\pi k}{n}\sinh2b.
\eea
The $\de_k$ are smooth functions of $\abar$ and $b$.
For $k=0$, (\ref{dek1}) and (\ref{dek2}) give
\bea
\cos\de_0\sinh\ga_0&=&\sinh2(\abar-b),
\\
\sin\de_0\sinh\ga_0&=&0,
\eea
so that with (\ref{ga0}) we have
\beq
\de_0=0,
\eeq
while from (\ref{gaksym}), (\ref{dek1}) and (\ref{dek2}) follows
for $k=1,\ldots,n$
\beq
\de_{2n-k}=-\de_k,
\eeq
implying
\beq
\de_n=0.
\eeq

The normalized eigenvectors of $M_k$ may now be written as
\beq
u_k^\ua=\frac{1}{\sqrt{2}}
\left(\ba{c}e^{\frac{i}{2}\de_k}
\\-ie^{-\frac{i}{2}\de_k}\ea\right),~~~~~~
u_k^\da=\frac{1}{\sqrt{2}}
\left(\ba{c}-ie^{\frac{i}{2}\de_k}
\\e^{-\frac{i}{2}\de_k}\ea\right),
\eeq
as may be checked by inserting them into (\ref{evecs}).
The eigenvectors $\psi_k^\pm$ of $V^\pm$, defined according to
(\ref{psi}) and (\ref{zk}) by
\beq
\psi_k^\uada=\frac{1}{\sqrt{n}}\left(\ba{c}e^{ik\pi/n}u_k^\uada\\
e^{2ik\pi/n}u_k^\uada\\\vdots\\e^{nik\pi/n}u_k^\uada\ea\right)
\eeq
are all orthogonal.
For different eigenvalues this is clear, while for the degenerate
eigenvalues $\ga_k$ and $\ga_{2n-k}$, $k=1,\ldots,n-1$, this is easily
seen by explicitly computing the scalar product of $\psi_k^\uada$ and
$\psi_{2n-k}^\uada$,
\bea
\psi_k^{\uada\dagger}\psi_{2n-k}^\uada
&=&
\frac{1}{n}u_k^{\uada\dagger}u_{2n-k}^\uada
\sum_{l=1}^n\left(e^{-ik\pi/n}e^{i(2n-k)\pi/n}\right)^l
\nn\\
&=&
\frac{1}{n}u_k^{\uada\dagger}u_{2n-k}^\uada
\left(\frac{1-e^{-2\pi i(n+1)k/n}}{1-e^{-2\pi ik/n}}-1\right)
\nn\\
&=&
0.
\eea

From here on, we will be considerably more explicit than \cite{kaufman}
(the following discussion is omitted altogether in \cite{huang}).
This is because, without the following steps, the solution is incomplete.

Let us define
\beq
R_+^{-1}=\left(\psi_1^\ua,\psi_{2n-1}^\da,\psi_3^\ua,\psi_{2n-3}^\da,
\ldots,\psi_{2n-3}^\ua,\psi_3^\da,\psi_{2n-1}^\ua,\psi_1^\da\right)
\eeq
and
$$
\hspace{-212pt}R_-^{-1}=\vspace{-20pt}
$$
\beq
\left(\psi_0^\ua,\psi_0^\da,\psi_2^\ua,\psi_{2n-2}^\da,
\psi_4^\ua,\psi_{2n-4}^\da,\ldots,\psi_{2n-4}^\ua,\psi_4^\da,
\psi_{2n-2}^\ua,\psi_2^\da\right)\!.
\eeq
The $R_\pm^{-1}$ are then unitary and diagonalize $V^\pm$
according to
\bea
\label{vpdiag}
\lefteqn{\!\!\!\!\!\!\!\!\!\!R_+V^+R_+^{-1}=}
\nn\\
&&\!\!\!\!\!\!\!\!\!\!\!\!\!\!\!\!
\left(\ba{ccccccc}
e^{+\ga_1}\\&e^{-\ga_1}\\
&&e^{+\ga_3}\\&&&e^{-\ga_3}\\
&&&&\ddots\\&&&&&e^{+\ga_{2n-1}}\\&&&&&&e^{-\ga_{2n-1}}
\ea\right)
\eea
and
\bea
\label{vmdiag}
\lefteqn{\!\!\!\!\!\!\!\!\!\!\!R_-V^-R_-^{-1}=}
\nn\\
&&\!\!\!\!\!\!\!\!\!\!\!\!\!\!\!\!\!
\left(\ba{ccccccc}
e^{+\ga_0}\\&e^{-\ga_0}\\
&&e^{+\ga_2}\\&&&e^{-\ga_2}\\
&&&&\ddots\\&&&&&e^{+\ga_{2n-2}}\\&&&&&&e^{-\ga_{2n-2}}
\ea\right)\!,
\eea
where (\ref{gaksym}) has been used.

The diagonal form above is not the most useful for the following
considerations.
Let us instead apply a further similarity transformation to obtain
again a form that is part of SO$(N,C)$.
First, there is a more useful way of writing $R_\pm^{-1}$.
For this purpose, define the $2\times2$-matrices
\bea
\lefteqn{D_{kl}
=(z_l^ku_l^\ua,z_{2n-l}^ku_{2n-l}^\da)}
\nn\\
&=&
\frac{1}{\sqrt{2}}
\left(\ba{rrr}
\exp i\left(\frac{kl\pi}{n}+\frac{\de_l}{2}\right)
&&-i\exp i\left(-\frac{kl\pi}{n}-\frac{\de_l}{2}\right)
\\\\
-i\exp i\left(\frac{kl\pi}{n}-\frac{\de_l}{2}\right)
&&\exp i\left(-\frac{kl\pi}{n}+\frac{\de_l}{2}\right)
\ea\right)
\nn\\
\eea
and organize them as
\beq
R_+^{-1}=\frac{1}{\sqrt{n}}\left(\ba{cccc}
D_{11}&D_{31}&\cdots&D_{2n-1,1}\\
D_{12}&D_{32}&\cdots&D_{2n-1,2}\\
\vdots&\vdots&\ddots&\vdots\\
D_{1n}&D_{3n}&\cdots&D_{2n-1,n}
\ea\right)
\eeq
and
\beq
R_-^{-1}=\frac{1}{\sqrt{n}}\left(\ba{cccc}
D_{01}&D_{21}&\cdots&D_{2n-2,1}\\
D_{02}&D_{22}&\cdots&D_{2n-2,2}\\
\vdots&\vdots&\ddots&\vdots\\
D_{0n}&D_{2n}&\cdots&D_{2n-2,n}
\ea\right).
\eeq
Now, define $R_X$ and its inverse by
\bea
\lefteqn{R_X^{\pm1}=}
\nn\\
&&\!\!\!\!
\left(\ba{cccc}
\exp(\mp i\pi\si_x/4)\\
&\exp(\mp i\pi\si_x/4)\\
&&\ddots\\
&&&\exp(\mp i\pi\si_x/4)\\
\ea
\right)
\nn\\
\eea
with
\beq
\exp(\mp i\pi\si_x/4)=\frac{1}{\sqrt{2}}
\left(\ba{rr}1&\mp i\\\mp i&1\ea\right)
\eeq
and apply the further similarity transformation
\beq
\label{simtrans}
V_S^\pm=R_XR_\pm V^\pm R_\pm^{-1}R_X^{-1}
\equiv S_\pm V^\pm S_\pm^{-1}
\eeq
to (\ref{vpdiag}) and (\ref{vmdiag}).
\begin{widetext}
This gives on the one hand
\bea
\label{vs}
V_S^\pm
&=&
\left(\ba{cccc}\exp(-\ga_{\cd{1}{0}}\si_y)\\
&\exp(-\ga_{\cd{3}{2}}\si_y)\\
&&\ddots\\&&&\exp(-\ga_{\cd{2n-1}{2n-2}}\si_y)\ea\right)
=
\exp\left(
\ba{cccc}-\ga_{\cd{1}{0}}\si_y\\
&-\ga_{\cd{3}{2}}\si_y\\
&&\ddots\\&&&-\ga_{\cd{2n-1}{2n-2}}\si_y\ea\right)
\nn\\
&=&
\exp\left(\sum_{\nu=1}^n
\ga_{\cd{$2\nu{-}1$}{$2\nu{-}2$}}J_{2\nu,2\nu-1}\right),
\eea
\end{widetext}
while on the other hand we have
\beq
S_+^{-1}=R_+^{-1}R_X^{-1}=\frac{1}{\sqrt{n}}\left(\ba{cccc}
\tilde{D}_{11}&\tilde{D}_{13}&\cdots&\tilde{D}_{1,2n-1,}\\
\tilde{D}_{21}&\tilde{D}_{23}&\cdots&\tilde{D}_{2,2n-1}\\
\vdots&\vdots&\ddots&\vdots\\
\tilde{D}_{n1}&\tilde{D}_{n3}&\cdots&\tilde{D}_{n,2n-1}
\ea\right),
\eeq
\beq
S_-^{-1}=R_-^{-1}R_X^{-1}=\frac{1}{\sqrt{n}}\left(\ba{cccc}
\tilde{D}_{10}&\tilde{D}_{12}&\cdots&\tilde{D}_{1,2n-2}\\
\tilde{D}_{20}&\tilde{D}_{22}&\cdots&\tilde{D}_{2,2n-2}\\
\vdots&\vdots&\ddots&\vdots\\
\tilde{D}_{n0}&\tilde{D}_{n2}&\cdots&\tilde{D}_{n,2n-2}
\ea\right),
\eeq
with
\beq
\tilde{D}_{kl}=
\left(\ba{rrr}\cos\left(\frac{kl\pi}{n}+\frac{\de_l}{2}\right)&&
-\sin\left(\frac{kl\pi}{n}+\frac{\de_l}{2}\right)\\\\
\sin\left(\frac{kl\pi}{n}-\frac{\de_l}{2}\right)&&
\cos\left(\frac{kl\pi}{n}-\frac{\de_l}{2}\right)\ea\right).
\eeq
Let us show now that the $S_\pm$ are elements of SO$(2n,C)$.
Since $R_\pm$ and $R_X$ are unitary, the $S_\pm$ are also unitary.
Since the $S_\pm$ are also real, they are orthogonal.
We still need to show that $\det S_\pm=1$.
Since $S_\pm$ are orthogonal, we have
\beq
(\det S_\pm)^2=\det S_\pm\det S_\pm^T=\det S_\pm\det S_\pm^{-1}=1
\eeq
and the question is only if $\det S_\pm=+1$ or $\det S_\pm=-1$.
Since $\S_\pm$ are analytic in $b$, we may work with $b=0$ and therefore
$\de_k=0$ and then analytically continue to non-zero $b$, upon which
$\det S_\pm$ cannot change discontinuously and therefore not change at all. 
We have
\beq
\det S_\pm^{-1}=\det R_X^{-1}S_\pm^{-1}R_X=\det R_X^{-1}R_\pm^{-1},
\eeq
where for $\de_k=0$
\beq
R_X^{-1}R_+^{-1}=\frac{1}{\sqrt{n}}\left(\ba{cccc}
\bar{D}_{11}&\bar{D}_{13}&\cdots&\bar{D}_{1,2n-1}\\
\bar{D}_{21}&\bar{D}_{23}&\cdots&\bar{D}_{2,2n-1}\\
\vdots&\vdots&\ddots&\vdots\\
\bar{D}_{n1}&\bar{D}_{n3}&\cdots&\bar{D}_{n,2n-1}
\ea\right),
\eeq
\beq
R_X^{-1}R_-^{-1}=\frac{1}{\sqrt{n}}\left(\ba{cccc}
\bar{D}_{10}&\bar{D}_{12}&\cdots&\bar{D}_{1,2n-2}\\
\bar{D}_{20}&\bar{D}_{22}&\cdots&\bar{D}_{2,2n-2}\\
\vdots&\vdots&\ddots&\vdots\\
\bar{D}_{n0}&\bar{D}_{n2}&\cdots&\bar{D}_{n,2n-2}
\ea\right),
\eeq
with
\beq
\bar{D}_{kl}=\left(\ba{cc}\exp(+ikl\pi/n)\\&\exp(-ikl\pi/n)\ea\right).
\eeq
For the purpose of computing their determinants, we may reorganize
$R_X^{-1}R_\pm^{-1}$ into block matrices
\beq
\left(\ba{cc}R_\pm^+\\&R_\pm^-\ea\right),
\eeq
such that all $\exp(+ikl\pi/n)$ are in the $n\times n$ matrices $R_\pm^+$
and all $\exp(-ikl\pi/n)$ in the $n\times n$ matrices $R_\pm^-$,
keeping the ordering of the $\exp(+ikl\pi/n)$ among themselves and of
the $\exp(-ikl\pi/n)$ among themselves.
Such a reorganization involves equal numbers of exchanges of rows and
columns, respectively, and does therefore not change the determinant of
$R_X^{-1}R_\pm^{-1}$.
Also, we have $R_\pm^-=R_\pm^{+*}$ and therefore
\bea
\det S_\pm^{-1}
&=&
\det R_X^{-1}R_\pm^{-1}=\det R_\pm^+\det R_\pm^-
\nn\\
&=&
|\det R_\pm^+|^2>0,
\eea
which leaves only
\beq
\det S_\pm=1.
\eeq
This means that the matrices $S_\pm$ are elements of SO$(2n,C)$ and can be
represented as
\beq
S_\pm=\exp(ic_{\al\be}^\pm J_{\al\be})
\eeq
with certain unknown complex parameters $c_{\al\be}^\pm$.

Let us summarize the crucial result of this section:
Applying similarity transformations
$V_S^\pm\equiv S_\pm V^\pm S_\pm^{-1}$
with certain matrices
$S_\pm=\exp(ic_{\al\be}^\pm J_{\al\be})$
to
$V^\pm=V_{a/2}V_b^\pm V_{a/2}$
with
$V_{a/2}=\prod_{\nu=1}^n\exp(\abar J_{2\nu,2\nu-1})$
and
$V_b^\pm=\exp(\mp2b J_{1,2n})\prod_{\nu=1}^{n-1}\exp(2b J_{2\nu+1,2\nu})$,
we obtain
$V_S^\pm=\exp\left(\sum_{\nu=1}^n
\ga_{\cd{$2\nu{-}1$}{$2\nu{-}2$}}J_{2\nu,2\nu-1}\right)$
with $\ga_k$ as defined in (\ref{gak}) and the subsequent sign
convention.
This is what we will use in the next section to determine all eigenvalues
of $\V$ and therefore the partition function $Z(a,b)$.

\section{Diagonalization of $\V$}
%--------------------------------
\label{vdiag}
Now use the same parameters $c_{\al\be}^\pm$ from the last section
to define the $2^n\times2^n$-dimensional transformation matrix
\beq
\S=\S_+\S_-,~~~~~~\S_\pm=\exp(ic_{\al\be}^\pm\J_{\al\be}^\pm),
\eeq
and write
\beq
\V_\S=\S\V\S^{-1}=\S_+\V^+\S_+^{-1}\S_-\V^-\S_-^{-1}
\equiv\V_S^+\V_S^-.
\eeq
The factors
\beq
\V_S^\pm
=\exp(ic_{\al\be}^\pm\J_{\al\be}^\pm)\V_{a/2}^\pm\Vb^\pm\V_{a/2}^\pm
\exp(-ic_{\al\be}^\pm\J_{\al\be}^\pm)
\eeq
have the same structure as
\beq
V_S^\pm=\exp(ic_{\al\be}^\pm J_{\al\be})V_{a/2}V_b^\pm V_{a/2}
\exp(-ic_{\al\be}^\pm J_{\al\be})
\eeq
from the last section.
Now imagine using the Baker-Campbell-Hausdorff formula \cite{cbh}
\bea
\lefteqn{\exp(A)\exp(B)=\exp\bigg(A+B-\frac{1}{2}[B,A]}
\nn\\
&&
+\frac{1}{12}\{[A,[A,B]]+[B,[B,A]]\}
+\cdots\bigg)
\eea
to work out all products of exponentials in $\V_S^\pm$.
Since the $\J_{\al\be}^+$, $\J_{\al\be}^-$ and $J_{\al\be}$ obey identical
algebras, the result is
\beq
\V_\S^\pm=\exp\left(\sum_{\nu=1}^n
\ga_{\cd{$2\nu{-}1$}{$2\nu{-}2$}}\J_{2\nu,2\nu-1}^\pm\right),
\eeq
so that
\bea
\V_\S
&=&
\exp\left[\sum_{\nu=1}^n
(\ga_{2\nu-1}\J_{2\nu,2\nu-1}^++\ga_{2\nu-2}\J_{2\nu,2\nu-1}^-)\right]
\nn\\
&=&
\exp\bigg[\frac{1}{4}\sum_{\nu=1}^n\ga_{2\nu-1}(\unit+\U_\X)\X_\nu
\nn\\
&&\hspace{20pt}
+\frac{1}{4}\sum_{\nu=1}^n\ga_{2\nu-2}(\unit-\U_\X)\X_\nu\bigg]
\eea
with the same $\ga_k$ as defined in (\ref{gak}) and the subsequent sign
convention.

To diagonalize $\V_\S$, define another similarity transformation
\beq
\V_\Y=\R_\Y\V_\S\R_\Y^{-1}
\eeq
with $\R_\Y$ and its inverse given by
\beq
\R_\Y^{\pm1}=2^{-n/2}\prod_{\nu=1}^n(\unit\pm i\Y_\nu).
\eeq
Since
\beq
\R_\Y\X_\nu\R_\Y^{-1}=\Z_\nu,
\eeq
this transformation takes $\V_\S$ into
\bea
\V_\Y
&=&
\R_\Y\V_\S\R_\Y^{-1}
=\exp\bigg[\frac{1}{4}\sum_{\nu=1}^n\ga_{2\nu-1}(\unit+\U_\Z)\Z_\nu
\nn\\
&&\hspace{75pt}
+\frac{1}{4}\sum_{\nu=1}^n\ga_{2\nu-2}(\unit-\U_\Z)\Z_\nu\bigg]
\hspace{20pt}
\eea
with
\beq
\U_\Z=\Z_1\cdots\Z_n.
\eeq

The matrix $\V_\Y$ is diagonal, but we still have to determine its elements.
$\U_\Z$ is a diagonal matrix with elements $+1$ and $-1$ occurring in equal
numbers.
For each element holds: If an even (odd) number of $\Z_\nu$ provides a
factor $-1$, the matrix element of $\U_\Z$ is $+1$ ($-1$).
This means: (i) A matrix element of $(\unit+\U_\Z)/2$ is $1$ ($0$) if an
even (odd) number of $\Z_\nu$ provides a factor $-1$;
(ii) A matrix element of $(\unit-\U_\Z)/2$ is $1$ ($0$) if an odd (even)
number of $\Z_\nu$ provides a factor $-1$.
It follows that the $2^n$ eigenvalues of $\V$ split into $2^{n-1}$
eigenvalues of the form
\beq
\label{oddevs}
\exp\left(\frac{1}{2}\sum_{\nu=1}^n(\pm)\ga_{2\nu-1}\right),
\eeq
and $2^{n-1}$ eigenvalues of the form
\beq
\label{evenevs}
\exp\left(\frac{1}{2}\sum_{\nu=1}^n(\pm)\ga_{2\nu-2}\right),
\eeq
where in the first (second) case all sign combinations with an even
(odd) number of minus signs occur.
This is reflected by the indices ``e'' and ``o'' in our result for the
partition function,
\begin{widetext}
\vspace{-0.65cm}
\bea
\lefteqn{Z(a,b)
=[2\sinh(2a)]^{mn/2}
\left[\sum_{\rm e}\exp\left(\frac{m}{2}
\sum_{\nu=1}^n(\pm)\ga_{2\nu-1}\right)
+\sum_{\rm o}\exp\left(\frac{m}{2}
\sum_{\nu=1}^n(\pm)\ga_{2\nu-2}\right)\right]}
\nn\\
&=&
\frac{1}{2}[2\sinh(2a)]^{mn/2}
\nn\\
&&\times
\left\{
\prod_{k=1}^n\left[2\cosh\left(\frac{m}{2}\ga_{2k-1}\right)\right]
+\prod_{k=1}^n\left[2\sinh\left(\frac{m}{2}\ga_{2k-1}\right)\right]
+\prod_{k=1}^n\left[2\cosh\left(\frac{m}{2}\ga_{2k-2}\right)\right]
-\prod_{k=1}^n\left[2\sinh\left(\frac{m}{2}\ga_{2k-2}\right)\right]
\right\}.
\nn\\
\eea
\end{widetext}
The last term within the braces has a sign differing from that in
\cite{kaufman}.
This is due to our different sign convention for $\ga_0$.
The eigenvalues of $\T$ are of course obtained by multiplying
(\ref{oddevs}) and (\ref{evenevs}) with the trivial factor
$[2\sinh(2a)]^{n/2}$.

The results for the eigenvalues of $\T$ and the partition function are
the starting point for the analysis of the thermodynamic properties of
the two-dimensional Ising model, the most interesting case being the
thermodynamic limit $m,n\rightarrow\infty$.
Such analyses can now proceed as usual (see e.g.\
\cite{onsager,kaufman,huang}) and will not be repeated here.

\section{Remarks About Magnetic Field and Three Dimensions}
%----------------------------------------------------------
\label{extensions}
Let us briefly remark on the difficulties encountered when trying to solve
the two-dimensional model with magnetic field or the three-dimensional
model in our approach.
In the first case, we need besides $\V_a$ and $\V_b$ another matrix
\beq
\Vc=\prod_{\nu=1}^n\exp(c\Z_\nu),
\eeq
and in the second case another matrix
\beq
\Vd=\prod_{\nu=1}^{n}\exp(d\Z_\nu\Z_{\nu+n'}),
\eeq
if the three-dimensional model is obtained from the two-dimensional one
by letting sites $\nu$ and $\nu+n'$ for $\nu=1,\ldots,n$ interact with
each other (assuming we have chosen $n$ such it can be divided by $n'$;
the lattice is then $m\times n'\times n/n'$ and has in the $n'$ and $n/n'$
directions the character of a screw on a torus, while in the $m$
direction it remains periodic).
For the three-dimensional model with magnetic field, we need both
$\Vc$ and $\Vd$.

Both the $\Z_\nu$ and the $\Z_\nu\Z_{\nu+n'}$ (with $n'>1$) are not
part of the algebra of $\J_{\al\be}^+$ and $\J_{\al\be}^-$.
In fact, it turns out that amending the Lie algebra with either of these
classes of matrices makes the number of elements in the new algebra grow
proportional to $2^n$ instead of $n^2$.
This eliminates the possibility of using an extended version of the
algebra of matrices $J_{\al\be}$ with side length growing proportional
to $n$.
But this was essential for the convenient periodicity properties of the
matrices $V^\pm$.

\end{document}